\documentclass[aps,preprint]{revtex4}%
\usepackage{amsfonts}
\usepackage{amsmath}
\usepackage{amssymb}
\usepackage{graphicx}%
\begin{document}
\preprint{ }
\title[Physical gauge]{Physical restrictions on the choice of electromagnetic gauge and their
practical consequences}
\author{H. R. Reiss}
\affiliation{Max Born Institute, 12489 Berlin, Germany}
\affiliation{American University, Washington, DC 20016-8058, USA}
\email{reiss@american.edu}

\pacs{32.80.Rm, 33.80.Rv, 42.50.Hz, 03.50.De}

\begin{abstract}
It is shown that electromagnetic potentials convey physical information beyond
that supplied by electric and magnetic fields alone, and are thus more
fundamental. Observable physical properties can impose conditions on the
selection of electromagnetic gauge (i.e. sets of potentials) that are explicit
and restrictive. This is true both classically and quantum mechanically. The
implication that the choice of gauge carries physical information is confirmed
by exhibiting a set of potentials that describes fields correctly, but that
violates physical constraints. The basic conclusions are that physical
requirements place limits on acceptable gauges; and that potentials are more
fundamental than fields in both classical and quantum physics, representing a
major generalization of the quantum-only Aharonov-Bohm effect. These important
properties are obscured if the dipole approximation is employed. The
properties demonstrated here relate directly to conditions that exist in
strong-field laser applications.

\end{abstract}
\date[17 February 2017]{}
\maketitle

\section{Introduction}

Electric and magnetic fields of electrodynamics can be represented by scalar
and vector potentials. Particular derivatives of these potentials will
generate the fields. These potentials are not unique, with alternative sets of
potentials connected by a mathematical procedure called a \textit{gauge
transformation}. The conventional point of view is that physical processes
depend only on the fields, and the potentials can be regarded as nothing more
than useful auxiliary quantities \cite{jackson}. An important exception to
this rule is the Aharonov-Bohm effect \cite{siday,ab}, in which a charged
particle passing near a solenoid containing a magnetic field can be deflected
by the potential that exists outside the solenoid in a field-free region. The
effect is particular to quantum mechanics, and provides only a cautionary
limitation to the general notion that fields are more basic than potentials.

It is shown here that such a ubiquitous phenomenon as a laser field imposes
strong limitations on possible gauge transformations when the demand is made
that the propagation property of the laser field must be sustained. This
limitation applies both classically and quantum mechanically. It is further
shown that if the laser field (a transverse field) impinges on a charged
particle that is simultaneously subjected to a Coulomb binding potential (a
longitudinal field), then the unique allowable gauge is the radiation gauge
(also known as the Coulomb gauge) if all physical constraints are to be
satisfied. This combination of transverse and longitudinal fields pervades
Atomic, Molecular, and Optical (AMO) physics. The common AMO practice of using
the dipole approximation in the description of laser-induced effects amounts
to replacing the transverse field of a laser by the fundamentally different
longitudinal field, thus altering basic physical constraints.

A brief review of the basic features of gauge transformations is given in the
next Section. The standard requirement is that the generating function for the
gauge transformation can be any scalar function that satisfies the homogeneous
wave equation \cite{jackson}. Section III considers the important case of a
propagating field. This application applies to all transverse fields,
including laser fields. It is shown that the only possible departure from the
familiar radiation gauge must be such that the 4-vector potential describing
the field can have added to it only a contribution that depends on the
light-cone coordinates appropriate to that field. This is an important
supplement to the standard gauge requirements of electrodynamics.

Section IV examines the more restrictive case where a charged particle is
simultaneously subjected to a transverse field (like a laser field) and a
longitudinal field (like a Coulomb binding potential). Since this combination
of fields describes most AMO situations, the basic electrodynamic principles
established here are directly applicable to strong-field laser experiments.
The operative limitation in this case comes from the properties of the
relativistic quantum equations of motion and the inferences that persist in
the nonrelativistic limit. These considerations are not important for laser
fields in the perturbative domain, but they are applicable under the
conditions that exist when fields are strong enough to require nonperturbative
methods. The reason is that laser fields propagate at the speed of light,
thereby introducing relativistic considerations even into nominally
nonrelativistic problems \cite{hr90}. Relativistic conditions exist and must
be properly accounted for in strong-field applications, since they signal the
importance of the magnetic field in ways that are invisible within the dipole approximation.

Section V exhibits the general conclusion that potentials are more fundamental
than the fields that are derived from them, by exhibiting two sets of
potentials that describe exactly the same electric and magnetic fields; but
one set satisfies all physical requirements, and the other set gives incorrect
predictions for such basic matters as the propagation property, Lorentz
symmetries, and the ponderomotive potential of a charged particle in the
field. This marks a major generalization of the important Aharonov-Bohm
effect, presently the sole practical example of the dominance of potentials
over fields.

The final Section is an overview of the essential results, including an
appraisal of the practical consequences of the results arrived at here. A
simple summary is that \textquotedblleft physical intuition\textquotedblright%
\ or \textquotedblright physical interpretation\textquotedblright\ is
dependent on the choice of electromagnetic gauge. The use of the dipole
approximation severely limits those benefits and can lead to the adoption of
physical pictures that do not match laboratory reality. A leading example is
the \textquotedblleft tunneling limit\textquotedblright\ that envisions a very
low frequency laser field as a nearly static electric field, in contrast to
the actuality of a laser field that propagates at the velocity of light for
all frequencies, and cannot possibly have a static limit. It is emphasized
that some seriously misdirected criteria have been adopted, unchallenged, in
strong-field physics.

\section{Basic Gauge Transformation}

For notational simplicity only vacuum conditions are considered, and Gaussian
units are employed. The electric field $\mathbf{E}$ and the magnetic field
$\mathbf{B}$ can be represented by the scalar potential $\phi$ and the vector
potential $\mathbf{A}$ as%
\begin{align}
\mathbf{E}  &  =-\mathbf{\nabla}\phi-\frac{1}{c}\partial_{t}\mathbf{A}%
,\label{a}\\
\mathbf{B}  &  =\mathbf{\nabla}\times\mathbf{A}. \label{b}%
\end{align}
A gauge transformation of the $\phi,$ $\mathbf{A}$ potentials to a new set
$\widetilde{\phi},$ $\widetilde{\mathbf{A}}$ can be achieved with the scalar
generating function $\Lambda$ with the connection that%
\begin{align}
\widetilde{\phi}  &  =\phi+\frac{1}{c}\partial_{t}\Lambda,\label{c}\\
\widetilde{\mathbf{A}}  &  =\mathbf{A}-\mathbf{\nabla}\Lambda. \label{d}%
\end{align}
The substitution of Eqs. (\ref{c}) and (\ref{d}) into (\ref{a}) and (\ref{b})
leaves the field expressions unchanged. The only constraint on $\Lambda$ is
that it should satisfy the homogeneous wave equation. This is to enable a
decoupling of the equations for the scalar and vector potentials. Relativistic
notation is useful. The 4-vector potential that encompasses both the scalar
and 3-vector potentials is%
\begin{equation}
A^{\mu}:\left(  \phi,\mathbf{A}\right)  , \label{e}%
\end{equation}
and the basic spacetime 4-vector is%
\begin{equation}
x^{\mu}:\left(  ct,\mathbf{r}\right)  . \label{f}%
\end{equation}
The expressions (\ref{c}) and (\ref{d}) are then subsumed into the single
expression%
\begin{equation}
\widetilde{A}^{\mu}=A^{\mu}+\partial^{\mu}\Lambda\label{g}%
\end{equation}
subject to the constraint on $\Lambda$ that it must satisfy%
\begin{equation}
\partial^{\mu}\partial_{\mu}\Lambda=0. \label{h}%
\end{equation}

\section{Gauge Limitation for Transverse Fields}

Now the important example is considered wherein $A^{\mu}$ represents a
\textit{transverse field}, with the equivalent terminologies that it is a
\textit{propagating field} or a \textit{plane-wave field}. Such a field
propagates in vacuum with the speed of light $c$, with the additional proviso
of special relativity that this speed of propagation must be the same in all
inertial frames of reference. This is equivalent to the statement (see, for
example, Refs. \cite{schwinger} and \cite{ss}) that any occurrence of $x^{\nu
}$ in the potential $A^{\mu}$ can only be in the form of the scalar product
with the propagation 4-vector $k^{\nu}$:%
\begin{equation}
\varphi\equiv k^{\mu}x_{\mu}, \label{i}%
\end{equation}
where%
\begin{align}
k^{\mu}  &  :\left(  \frac{\omega}{c},\mathbf{k}\right)  ,\label{j}\\
\left\vert \mathbf{k}\right\vert  &  =\omega/c. \label{k}%
\end{align}
In other words, $k^{\mu}$ is a lightlike 4-vector with the 3-vector component
$\mathbf{k}$ in the propagation direction of the transverse field. The
4-vector potential $A^{\mu}$ can depend on the spacetime 4-vector $x^{\mu}$
only as $A^{\mu}\left(  \varphi\right)  $, and the same condition must apply
to $\widetilde{A}^{\mu}.$ Therefore, Eq. (\ref{g}) requires that
$\partial^{\mu}\Lambda$ must also be a function of $\varphi$ alone. This
requirement can be satisfied by the condition that $\Lambda$ depends on
$x^{\mu}$ only as $\Lambda=\Lambda\left(  \varphi\right)  $, since then%
\begin{equation}
\partial^{\mu}\Lambda\left(  \varphi\right)  =\partial^{\mu}\left(
\varphi\right)  \frac{d}{d\varphi}\Lambda\left(  \varphi\right)  =k^{\mu
}\Lambda^{\prime}\left(  \varphi\right)  , \label{l}%
\end{equation}
where $\Lambda^{\prime}$ is the total derivative of $\Lambda$ with respect to
$\varphi$. The gauge-transformed 4-vector potential $\widetilde{A}^{\mu}$ can
therefore differ from the original $A^{\mu}$ only by a quantity that lies on
the light cone, since the gauge-transformation condition (\ref{g}) must be of
the form%
\begin{equation}
\widetilde{A}^{\mu}=A^{\mu}+k^{\mu}\Lambda^{\prime}. \label{m}%
\end{equation}

The condition (\ref{m}) is very restrictive. One important consequence is that
the squared 4-vector potential is gauge invariant \cite{hrmod,hrpond}, as
follows from the light-cone condition%
\begin{equation}
k^{\mu}k_{\mu}=0, \label{n}%
\end{equation}
and the transversality condition%
\begin{equation}
k^{\mu}A_{\mu}=0, \label{o}%
\end{equation}
so that Eq. (\ref{m}) leads to%
\begin{equation}
\widetilde{A}^{\mu}\widetilde{A}_{\mu}=A^{\mu}A_{\mu}. \label{p}%
\end{equation}
The ponderomotive potential of a charged particle in a transverse field is
proportional to $A^{\mu}A_{\mu}$, meaning that this fundamentally important
quantity \cite{hrmod,hrpond} is gauge-invariant.

An important \textit{caveat} is that the use of the dipole approximation, a
standard procedure in AMO physics, has the effect of losing altogether the
propagation property of a laser field. The dipole approximation amounts to the
replacement of the propagating, transverse field by a simple oscillatory
electric field, with the significance that the basic condition of Eq.
(\ref{m}) is discarded. That is, Eq. (\ref{g}) no longer leads to Eq.
(\ref{m}) when the dipole approximation is employed.

It is well-known from long experience in nuclear and high energy physics that
calculations of the effects of plane wave fields on charged particles can be
successfully applied in the context of the radiation gauge. A convenient way
to describe the radiation gauge is that it is the gauge within which a pure
transverse field is described by the 3-vector potential $\mathbf{A}$ alone,
and a pure longitudinal field is described by a scalar potential $\phi$ (or
$A^{0}$) alone.

\section{Combined Transverse and Longitudinal Fields}

The ionization of an atom by a laser field typifies AMO processes. An atomic
electron in a laser field experiences both the transverse field of the laser
and the longitudinal field of the binding potential. The dipole approximation
has been useful in AMO physics since it offers the simplifying property that
the laser field is replaced by an oscillatory electric field, thus
substituting another longitudinal field for the transverse field. That is,
traditional AMO physics replaces a combination of a transverse laser field and
a longitudinal binding potential by two longitudinal fields. This is true
whether the so-called \textquotedblleft length gauge\textquotedblright\ is
used, where the interaction Hamiltonian is of the form $\mathbf{r\cdot
E}\left(  t\right)  $ or by the gauge-equivalent \cite{gm} \textquotedblleft
velocity gauge\textquotedblright\ where the interaction Hamiltonian contains
$\mathbf{A}\left(  t\right)  \mathbf{\cdot p}$.

When laser fields are very strong, the fact that the field propagates with the
velocity of light becomes an important feature \cite{hr90}. Replacement of the
propagating field by the dipole-equivalent oscillatory electric field is no
longer sufficient. Even when magnetic forces remain small, the complete
neglect of the magnetic field removes all possibility of propagation. This has
major importance in practical applications. For example, the laboratory
detection of \textquotedblleft Above-Threshold Ionization\textquotedblright%
\ (ATI) in 1979 \cite{ati}, where the perturbation-theory dominance of the
lowest-order process gives way to the participation of higher orders of
interaction with the applied field, caused a major sensation in the AMO
community and triggered theoretical efforts lasting more than a decade (see
the introductory remarks in Ref. \cite{cbc1}) to achieve some understanding of
how this could happen in a dipole-approximation context. By contrast, a theory
based on the nonrelativistic limit of a relativistically formulated theory
\cite{hr90} provided an anticipatory prediction of all of the ATI features
\cite{hr80} in a theory paper prepared in advance of the observation of ATI.
It is important to note that a nonrelativistic limit of a relativistic theory
leads to analytical forms that resemble theories based on \textit{a priori}
employment of the dipole approximation, but the seemingly slight differences
are nevertheless critical.

A laser-induced phenomenon that is unquestionably relativistic is the
production of electron-positron pairs. It was predicted in 1971 \cite{hrpairs}
and confirmed in 1997 \cite{burke} that this was possible with a laser
wavelength of the order of $1\mu m$ at a focused field intensity of about
$10^{18}W/cm^{2}$. Many laboratories can now produce such intensities; it is
simply a confirmation of the need to recognize the relativistic foundations of
laser effects.

Reduction of a relativistic treatment to a nonrelativistic limit of the
effects of combined transverse and longitudinal fields introduces a feature
that had not been anticipated. Following the usual practice of neglecting
effects of the spin of the electron, the universally employed relativistic
description of the electron is the Klein-Gordon (KG) equation
\cite{pauliweiss,schweber}%
\begin{equation}
\left[  \left(  i\hbar\partial_{\mu}-\frac{q}{c}A_{\mu}\right)  \left(
i\hbar\partial^{\mu}-\frac{q}{c}A^{\mu}\right)  -m^{2}c^{2}\right]  \psi=0.
\label{p1}%
\end{equation}
A separation of time and space parts gives the form%
\begin{equation}
\left[  \left(  \frac{i\hbar}{c}\partial_{t}-\frac{q}{c}A^{0}\right)
^{2}-\left(  -i\hbar\mathbf{\nabla}-\frac{q}{c}\mathbf{A}\right)  ^{2}%
-m^{2}c^{2}\right]  \psi=0. \label{p2}%
\end{equation}
It is the first parenthesis in the square bracket of Eq.(\ref{p2}) that is of
interest here, since $A^{0}$ can represent the time part of the laser-field
4-vector potential $A_{PW}^{0}$ (where the subscript $PW$ stands for
\textquotedblleft plane wave\textquotedblright) as well as a binding potential
$V$ that may be present:%
\begin{equation}
A^{0}=A_{PW}^{0}+V. \label{p3}%
\end{equation}
When $V$ is a Coulomb binding potential,%
\begin{equation}
V\sim1/r, \label{p4}%
\end{equation}
this singularity causes problems \cite{schoene} in the reduction of the KG
equation to the Schr\"{o}dinger equation in the nonrelativistic limit
\cite{schiff}. The act of squaring indicated for the first term in
Eq.(\ref{p2}) introduces a cross coupling $VA_{PW}^{0}$ that is also singular
at the origin of spatial coordinates, but it is an unacceptable term because
the magnitude of this singular term depends upon the properties of the laser
field. This unphysical behavior will not occur if%
\begin{equation}
A_{PW}^{0}=0, \label{p5}%
\end{equation}
which corresponds to selection of the radiation gauge (also known as the
Coulomb gauge) wherein longitudinal fields are represented by scalar
potentials and transverse fields are represented by 3-vector potentials. If
this gauge selection must be enforced in a relativistic problem, that gauge
must also refer to the nonrelativistic limit.

The same considerations arising in the reduction of the KG equation to the
Schr\"{o}dinger equation in the nonrelativistic limit applies also to the
reduction of the Dirac equation to the Pauli equation for a spin-%
%TCIMACRO{\U{bd} }%
%BeginExpansion
${\frac12}$
%EndExpansion
particle. The is most easily seen from the second-order form of the Dirac
equation \cite{feyngell,schweber}:%
\begin{gather}
\left[  \left(  i\hbar\partial_{\mu}-\frac{e}{c}A_{\mu}\right)  \left(
i\hbar\partial^{\mu}-\frac{e}{c}A^{\mu}\right)  +\frac{1}{2}\frac{e\hbar}%
{c}\sigma^{\mu\nu}F_{\mu\nu}-m^{2}c^{2}\right]  \psi=0,\label{q1}\\
\sigma^{\mu\nu}=\frac{1}{2i}\left(  \gamma^{\mu}\gamma^{\nu}-\gamma^{\nu
}\gamma^{\mu}\right)  , \label{q2}%
\end{gather}
where the $\gamma^{\mu}$ are the standard Dirac matrices and $F^{\mu\nu}$ is
the electromagnetic field tensor. This equation for a spin-%
%TCIMACRO{\U{bd} }%
%BeginExpansion
${\frac12}$
%EndExpansion
particle is the same as the KG equation (\ref{p1}) for a spin-0 particle,
but\ with the addition of a term representing the spin interaction. It
presents the same dilemma in reduction to the Pauli equation as does the KG
equation in reduction to the Schr\"{o}dinger equation. The second-order Dirac
equation has the dual advantages of its similarity to the KG equation, as well
as lacking the \textit{Zitterbewegung }problem \cite{schweber,bd} of the first
order Dirac equation in reduction to the nonrelativistic limit.

Elimination of the unphysical coupling of the laser field to a singular
quantity means that
\begin{equation}
A_{PW}^{0}=0, \label{r}%
\end{equation}
a feature of the radiation gauge, is a general requirement for attainment of
the correct equation of motion. Explicitly, since Eq. (\ref{m}) is a general
requirement for a propagating field, and it has now been shown that the
additional presence of a binding potential requires that any $\widetilde
{A}^{0}$ must also vanish, the condition (\ref{m}) means that
\begin{equation}
\Lambda^{\prime}=0,\quad\Lambda=\text{ constant} \label{s}%
\end{equation}
must hold true. That is, no departure from the radiation gauge for the
transverse field is allowable if all field conditions are to be satisfied exactly.

None of the above reasoning arises if the dipole approximation is imposed from
the outset. This does not mean that the dipole approximation is more
convenient because of this; it means rather that the dipole approximation
infers a measure of approximation beyond the usual interpretation. This is not
consequential for fields that are perturbatively weak, but it is of
fundamental importance when applied transverse fields are strong. A practical
example of this has already been mentioned: the ATI phenomenon is perplexing
within the dipole approximation \cite{cbc1}, but it is obvious when
propagating-field considerations are \textit{a priori} present
\cite{hr80,hr90}.

\section{Potentials Are More Fundamental Than Fields}

The strong constraints that have been found to apply to potentials, but
without reference to the fields associated with those potentials, has
immediate significance. That is, potentials have introduced essential probes
into physical phenomena such as the propagation phenomenon, preservation of
the ponderomotive energy, proper reduction to the Schr\"{o}dinger equation,
and so on. These properties become evident from the potentials, but not from
the fields.

A specific simple (but fundamental) example is now given where two sets of
potentials can be written for description of the same fields, where one set of
potentials is acceptable, but the other is nonphysical. A monochromatic
plane-wave field of constant amplitude can be described by the 4-potential%
\begin{equation}
A^{\mu}\left(  \varphi\right)  =A_{c}^{\mu}\cos\varphi, \label{t}%
\end{equation}
where the phase $\varphi$ is given in Eq. (\ref{i})%
\begin{equation}
\varphi=k^{\mu}x_{\mu}=\omega t-\mathbf{k\cdot r}, \label{u}%
\end{equation}
and $A_{c}^{\mu}$ is a constant 4-vector amplitude. It is noted here that this
4-potential satisfies the Lorenz condition%
\begin{equation}
\partial^{\mu}A_{\mu}=k^{\mu}A_{\mu}^{\prime}\left(  \varphi\right)  =0
\label{v}%
\end{equation}
because of the transversality condition of Eq. (\ref{o}). (The Danish
physicist L. V. Lorenz should not be confused with the Dutch physicist H. A. Lorentz.)

Now consider the gauge transformation generated by the function \cite{hr79}%
\begin{equation}
\Lambda=-A^{\mu}\left(  \varphi\right)  x_{\mu}. \label{w}%
\end{equation}
This gives the transformed 4-potential%
\begin{equation}
\widetilde{A}^{\mu}=-k^{\mu}\left(  x^{\nu}A_{\nu}^{\prime}\right)  .
\label{x}%
\end{equation}
It is readily verified that
\begin{equation}
\partial^{\mu}\partial_{\mu}\Lambda=0, \label{y}%
\end{equation}
the sole condition normally required of the generating function of a gauge
transformation \cite{jackson}. Because $\widetilde{A}^{\mu}$ was obtained from
the $A^{\mu}$ of Eq. (\ref{t}) by a gauge transformation, the electric and
magnetic fields obtained from (\ref{x}) are identical to those obtained from
(\ref{t}), as can be verified by direct computation. It is also true that
$\widetilde{A}^{\mu}$ is a Lorenz gauge, and it is even true that
$\widetilde{A}^{\mu}$ is transverse because of the light-cone condition
(\ref{n}).

However, the vector potential $\widetilde{A}^{\mu}$ given in Eq. (\ref{x}) is
not a physically acceptable gauge. It has the incorrect Lorentz transformation
property of being lightlike rather than spacelike. It predicts that the
all-important ponderomotive energy \cite{hrmod,hrpond} vanishes, because%
\begin{equation}
\widetilde{A}^{\mu}\widetilde{A}_{\mu}=0, \label{z}%
\end{equation}
and it does not possess the basic property required by relativity that it
depend on the spacetime 4-vector $x^{\mu}$ only in the combination $k^{\mu
}x_{\mu}$ as demanded by the condition (\ref{i}). All of these failures occur
for the simple reason that the gauge transformation (\ref{w}) that produced
$\widetilde{A}^{\mu}$ does not depend on $x^{\mu}$ solely in the form of the
scalar product (\ref{i}). Nevertheless, the unphysical nature of
$\widetilde{A}^{\mu}$ is not evident from the normal rules for performing a
gauge transformation. Judged by prediction of the correct electric and
magnetic fields, one would be justified in employing the $\widetilde{A}^{\mu}$
of (\ref{x}) as the gauge-equivalent version of (\ref{t}). However, this
seemingly safe conclusion based on the fields is incorrect because of the
unphysical properties that are evident only by noting that the physical
properties of the 4-potential (\ref{t}) are different from those of the
4-potential (\ref{x}).

It is not enough to know the fields; one must know the appropriate potentials.

\section{Practical Consequences}

The focus of attention throughout this article is on the properties of
propagating fields, with the specific case of laser fields as the most
important practical example. It has been shown that when a laser field
interacts with matter, so that bound charged particles are subjected
simultaneously to both transverse and longitudinal fields, then the only
formally acceptable electromagnetic gauge that can be employed is the
radiation gauge (also called the Coulomb gauge). It has been remarked that
this restriction is not of major importance when fields are perturbatively
weak, but there is a great and growing interest in the effects of very strong
laser fields. The practical models currently employed in strong field
applications are based on the dipole approximation, which amounts to treating
the laser field as an oscillatory electric field, with no propagation property
at all, and the results here obtained do not apply in the dipole context.
Since the dipole approximation gives the appearance of introducing important
conceptual and practical simplifications, and many successes have been
achieved in this manner, it is natural to inquire about the practical
consequences of the results shown above. That is a fundamental question, and a
comprehensive answer is proposed.

Since laser fields are, in actuality, transverse, propagating fields, it is to
be expected that physical understanding of practical consequences of laser
interactions with matter should be based on the properties of propagating
fields. The dipole approximation reduces the laser field to an oscillatory
electric field, which is a longitudinal field that differs fundamentally from
an actual laser field. One important example has already been mentioned. The
ATI phenomenon, so startling and unexpected within the AMO community, is
actually an obvious and commonplace consequence of all strong-field phenomena.
For example, in the context of pair production by strong laser fields, one
finds the 1971 comment \cite{hrpairs}: \textquotedblleft...an extremely high
order process can be competitive with -- and even dominate -- the lowest order
... process.\textquotedblright\ In the context of strong-field bound-bound
transitions, it was shown in 1970 that \cite{prl25}: \textquotedblleft...as
the intensity gets very high, ... the lowest order process gets less probable
... [and] higher-order processes become increasingly
important.\textquotedblright\ The 1980 strong-field approximation (SFA) paper
demonstrates the basic aspects of ATI, including some that were not observed
in the laboratory until much later. For example, the character of spectra
generated by strong, circularly polarized fields, exhibiting a multi-peaked
spectrum with a near-Gaussian envelope with the most probable order being
significantly higher than the lowest order, was observed with astonishment in
a 1986 experiment \cite{bucks86}, but this was already predicted in 1980, and
the 1980 theory was accurate in exhibiting \cite{hr87}\ the explicit behavior
found in the 1986 experiment. The reminder is important here that, although
the 1980 paper superficially resembles dipole-approximation theories, it is
actually the nonrelativistic limit of a relativistic theory of laser-induced
ionization \cite{hr90,hrrel}. The distinction is vital.

The above paragraph reveals that a propagating-field theory, since it models
the actual laser field, can produce results that are more insightful and more
successful than theories based on an oscillating-electric-field model.
Furthermore, the 1980 SFA theory is actually easier to apply than the
dipole-approximation versions of the SFA.

A recent example is instructive. In very precise spectrum measurements in an
ionization experiment with circularly polarized light, it was found to be
possible to detect the effects of radiation pressure on the photoelectrons
\cite{smeenk}. Attempts to provide a theoretical explanation for the effect in
a dipole-approximation context proved to be extremely difficult and
inconclusive \cite{smeenk,cbc1,cbc2}. This is not surprising. Radiation
pressure arises from photon momentum that does not exist in a
dipole-approximation theory. In the context of a transverse-field description,
the most probable kinetic energy of a photoelectron released by a strong,
circularly polarized field is just the ponderomotive energy $U_{p}$. The
number of photons above threshold needed to produce such a photoelectron is
$n=U_{p}/\hbar\omega$. Each photon carries a momentum of $\hbar\omega/c$, with
all photon momenta aligned in the direction of propagation of the laser field,
so the field-induced momentum in the propagation direction is just $U_{p}/c$,
and this is independent of the atom being ionized when the field is strong.
This is precisely what the laboratory measurements reveal
\cite{smeenk,hrpress}. Transverse-field concepts produce insightful and
quantitatively accurate results as shown in the span of three sentences given
above, as contrasted with three journal articles \cite{smeenk,cbc1,cbc2}.

The seeming simplicity of dipole-approximation methods is actually
counter-productive in strong fields, as shown by the ATI and radiation
pressure examples. The dipole approximation can lead to complication rather
than simplicity.

Perhaps the most consequential of all misconceptions that arise from
dependence on a dipole-approximation model of laser effects is the matter of
low frequency behavior \cite{hr101,hrtun}. The oscillatory electric fields
that arise from a dipole-approximation theory approach a constant electric
field as the frequency declines. This limit (sometimes called the
\textit{tunneling limit}) has been applied as a test of the accuracy of
theoretical models. For example, a textbook on the subject of strong
laser-field effects altogether rejects models based on transverse fields,
since they do not approach the tunneling limit \cite{jkp}. Another example is
a paper that assesses the accuracy of analytical approximations based on their
behavior as the field frequency approaches zero \cite{jarek}. However, actual
transverse fields in vacuum always propagate at the speed of light
independently of frequency. There is no limit possible in which a real
propagating field becomes a static field. The effect on strong-field theory of
this zero-frequency misconception continues to the present. It is related to
the equally problematic concept that the final arbiter of validity is to be
found in the exact numerical solution of the Schr\"{o}dinger equation,
generally referred to as TDSE (Time-Dependent Schr\"{o}dinger Equation). Since
TDSE as customarily employed is based on the dipole approximation, it
reinforces the critically misleading concept that there is a zero-frequency
limit of laser effects equivalent to a constant electric field.

It is difficult to conceive of a notion more consequential for an entire field
of inquiry than this reliance on the criterion that laser-induced effects have
a zero-frequency limit equivalent to that of a constant electric field. When
joined with the equally difficult concept, championed by K.-H. Yang
\cite{yang} and others \cite{lamb}, that the scalar potential known as the
length gauge (accurate for a scalar field like a longitudinal field) can
somehow be a privileged gauge for the description of a vector field like the
transverse field of a laser beam, the discipline of strong-field physics is
laboring under a burden of misdirected criteria. This has stood nearly
unchallenged since the 1979 observation of ATI \cite{ati}. The scrutiny
provided by an approach based on the radiation gauge creates the necessary challenge.

\end{document}